# The Discrete Logarithm Problem in the ElGamal Cryptosystem over the Abelian Group U(n) Where n= $p^m$, or $2p^m$

Hayder Raheem Hashim
Assistant Lecturer
Department of Mathematics
Faculty of Mathematics & Computer Science
University of Kufa, Iraq

*Abstract*— This study is mainly about the discrete logarithm problem in the ElGamal cryptosystem over the abelian group U(n) where n is one of the following forms $p^m$, or $2p^m$ where p is an odd large prime and m is a positive integer. It is another good way to deal with the ElGamal Cryptosystem using that abelian group  U(n)={x: x is a positive integer such that x<n and gcd(n,x)=1}  in the setting of the discrete logarithm problem . Since I show in this paper that this new study maintains equivalent ( or better) security with the original ElGamal cryptosystem( invented by Taher ElGamal in 1985)[1], that works over the finite cyclic group of the finite field. It gives a better security because theoretically ElGamal Cryptosystem with U($p^m$) or with U($2p^m$) is much more secure since the possible solutions for the discrete logarithm will be increased , and that would make this cryptosystem is hard to broken even with thousands of years.

*Keywords*— ElGamal Cryptosystem, The abelian group U(n), The Discrete Logarithm Problem over U(n), The ElGamal cryptosystem over U(n) : n =$p^m$, or $2p^m$ for a positive integer m and  p is an odd large prime.

I. Introduction

 The Classical ElGamal cryptosystem  works over the finite cyclic group of the finite field , and it's one of  the most popular and widely used cryptosystems. It's described in the setting of the multiplicative cyclic group  $Z^*_p$ where p is a large  prime the field Zp [2].  In this paper I modify the ElGamal Cryptosystem by using the abelian multiplicative group U(n) modulo a large integer n(where n  is one of the following forms $p^m$, or $2p^m$  for a positive integer m and  p is an odd  large prime) instead of $Z^*_p$ and get equivalent results. However it's proven that U(n) is a group[1] , I will use my own words to prove that for every n >1 , U(n) is an abelian group under the multiplication and use the proven theorem that  U(n) is a cyclic group where n  is one of the following forms $p^m$, or $2p^m$   for a positive integer m and  p is an odd  prime in this new study of this paper. Then I construct the public and private keys using the new setting of the  discrete logarithm problem , the encryption and decryption procedure  of the ElGamal cryptosystem using U(n), and finally illustrate example to show the procedure of the ElGamal cryptosystem using U(n). (Where  n  is one of the following forms $p^m$, or $2p^m$   for a positive integer m and  p is an odd  large prime) .

**A-*Theorem* :** For every n>1 , prove that the set U(n)={x : x is a positive integer such that x<n and gcd(n,x)=1} is an abelian group under the multiplication modulo n.
**Proof:** First of all, let's show that the multiplication mod n is a binary operation on U(n), i.e for all a and b in U(n), a*b is  in U(n). Assume for a contradiction that a*b is not  in U(n). suppose that gcd(a*b, n)=q where the integer q>1, then q\a*b and q\n.[6]
Therefore, (q\a or q\b) and q\n.
If q\a and q\n, then gcd(a,n)$\neq$1 , and similarly if q\b and q\n, then gcd(b,n)$\neq$1. That would contradicts that a and b are in U(n). Hence, a*b is in U(n).
To show that U(n) is a group,
-Associativity : Since the multiplication modulo n is associative, I'll assume that U(n) is associative .
-Identity : The identity element for the multiplication modulo n is 1 , and 1 is an element in U(n).
-Inverses : Suppose that a is in U(n), then gcd(n,a)=1. Therefore, there exists two integers w and v such that , aw+nv=1 . So, (aw+nv) mod n$\equiv$ 1 which leads to a*w mod n $\equiv$1, then w is the inverse of a .
BUT it is important to show that w is in U(n). Since wa+nv=1, then gcd(w,n)=1. Hence,  w is in U(n).
Therefore, U(m) is a group under the multiplication modulo m>1[3] .

Note that, the multiplication is commutative, so U(m) is an abelian group under the multiplication.

**B-Theorem[8]** : The group U(n) is cyclic if and only if n is of one of the following forms : 2, 4, $p^m$, $2p^m$ where m is a positive integer and p is an odd prime.

***C-Discrete Logarithm Problem over the abelian group ( U(n) : n= $p^m$,or $2p^m$ )***

If G is the finite abelian group U(n) (under the multiplication) and g is a generator of U(n) since U(n) is a cyclic group[4 ], then every element h in U(n) can be written as gx for some integer x. The discrete logarithm to the base g of h in the group U(n) is defined to be x .

*1- Example:* To illustrate discrete logarithm problem by a simple example, let n=p1=17, then U(17) is a cyclic( abelian) group generated by 3. So for any integer x , 3x (mod 17) belongs to U(17)={1,2,….,16}. If x=4 is chosen, then 34 mod 17≡ 13. But reversing the previous problem such that 3y ≡ 13 (mod 17), then finding integer y is the discrete logarithm problem. Therefore, the discrete logarithm problem is a one way function that is easy to perform but hard to reverse. I.e, for U(17)=<3>. We see that,

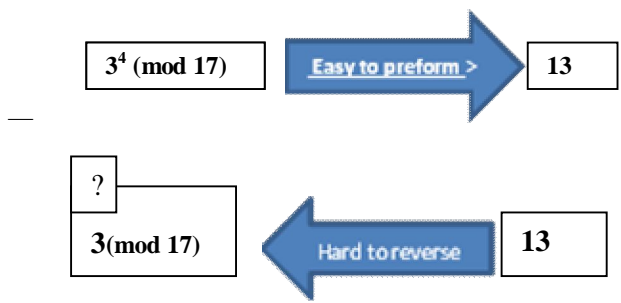

*2-Remark* [5]: In order to make n as a safe prime (large) number when it's using U(n) as the basis of discrete logarithm based cryptosystems, n should be chosen such that n is of the forms pm,or 2pm where m is positive and p is a very large prime number (usually at least 1024 –bit).

***D-The ElGamal Cryptosystem over( U(n): where n= $p^m$,or $2p^m$ )***

*1- Generating a public-key and private key of the EGamal Cryptosystem over ( U(n): where n= pm,or 2pm ):-*
{
-Select a large prime p.
– Select a group G=U(n) such that n= $p^m$, or n= $2p^m$.
-Select a random integer a to be a member of the group G=U(n) under the multiplication such that 1< a ≤n-2.
-Select r1 to be a generator of U(n).
-Call  r2 ≡( r1)$^a$ (mod n). [6]
-Let the public-key→(r1, r2, n)//(To be announced publicly). [6]-Let the private key →(a) // (To be kept secret ). [6]
}

*2-The ElGamal Encryption: (r1, r2, n, Pi). // Pi: the plaintext :[7]*
{
-Converting or translating the letters in the Plaintext into their numerical equivalents by using the table below (table 1) as example if the plaintext has just letters.
-Then for i=1,2,….,N, form blocks (Pi)of the largest possible size (with even number of digits).
-For each Pi, select a random integer ki with , 1≤ ki ≤n-2 (k could be the same for all the blocks).
- C1,i ≡r1ki (mod n).
-C2,i ≡ (Pi*$r_2^{ki}$) (mod n) .

*International Journal of Mathematics Trends and Technology – Volume 7  Number 3 – March 2014*

-The cipher corresponding to the plaintext block Pi is the ordered pair E(Pi)=(C1,i,C2,i).
-Return C1,i and C2,i. // C1,i and C2,i :The cipher-texts
}

*3-The ElGamal Decryption:(a,n, C1,i,C2,i).[6]*
{
-Pi≡ [C2,i(C1,ia)-1] (mod p) for all i=1,2,…..,N
-Return Pi.-
          }
Note that:
* $C_{1,i} \equiv r_1^{k_i}$ (mod n).
* $C_{2,i} \equiv (P_i * r_2^{k_i})$ (mod n) .
* $r_2 \equiv r_1^a$ (mod n) .

*4-Note:* To summarize The ElGamal Cryptosystem Procedure over( U(n) : n= pm,or 2pm   ) see figure 1 below .

*5-Example:* Let's consider a simple example to show how The ElGamal cryptosystem works over U(n) where   n= $p^m$, or 2$p^m$ ,for  an integer  m≥1,
 Suppose that "Sarah " wants to send a message" I like math "  to "Niwar" whose Public-key is (r1, r2, n=$p^1$) =(3, 23, 29) and whose private key a=4. (Note that $r_1$=3 is a generator of U(29)).

*5.1- What Sarah has to do to encrypt that message to be sent to Niwar is the following:*
1.1) Translate the message to its numerical equivalents(by using the table above), then group it to blocks with even number of digits (Will use two digits).
Therefore, the message : "I like math" becomes.
Plaintext , Pi: " 08 11 08 10 04 12 00 19 07 "
1.2)Select a random integer 1≤ ki≤ n-2 (Choose ki=5)  for all i=1,…..,8.
1.3) Use Niwar's Public-key , (r1, r2, n) =(3, 23, 29) with k=5 to encrypt each Plain-text block P in a cipher-text using the relationship : E(Pi)=( C1,i, C2,i) such that C1,i ≡$(r_1)^k$ (mod n) and C2,i ≡Pi*$(r_2)^k$ (mod n) which as the following :-
C1,i ≡$(r_1)^k$ (mod n)≡ 35 (mod 29)=11.
Then for all i, compute C2,i for each block Pi as in table 2 below.

5.2- What Niwar has to do when he gets the following Ciphertext Message :
E(Pi)=(C1,i,C2,i)= (11,26) (11,14) (11,26) (11,18) (11,13) (11,10) (11,0) (11,11) (11,1) . Then he wants to decrypt it, as the following:
2.1)Use his prime number n=p=29 and his private key a= 4, and the Plaintext formula : Pi ≡(C2,i)((C1,i)a)-1(mod 29) to decrypt each (C1,i,C2,i)  for all i=1,…..,9. (see table 3 below ).Therefore the plaintext in the numerical equivalents form is ,
                 Pi: " 08 11 08 10 04 12 00 19 07 "
2.2)Then by translating this back to letters, we obtain the right message back,

"I L I K E Math ".

*E-Figures and Tables:*

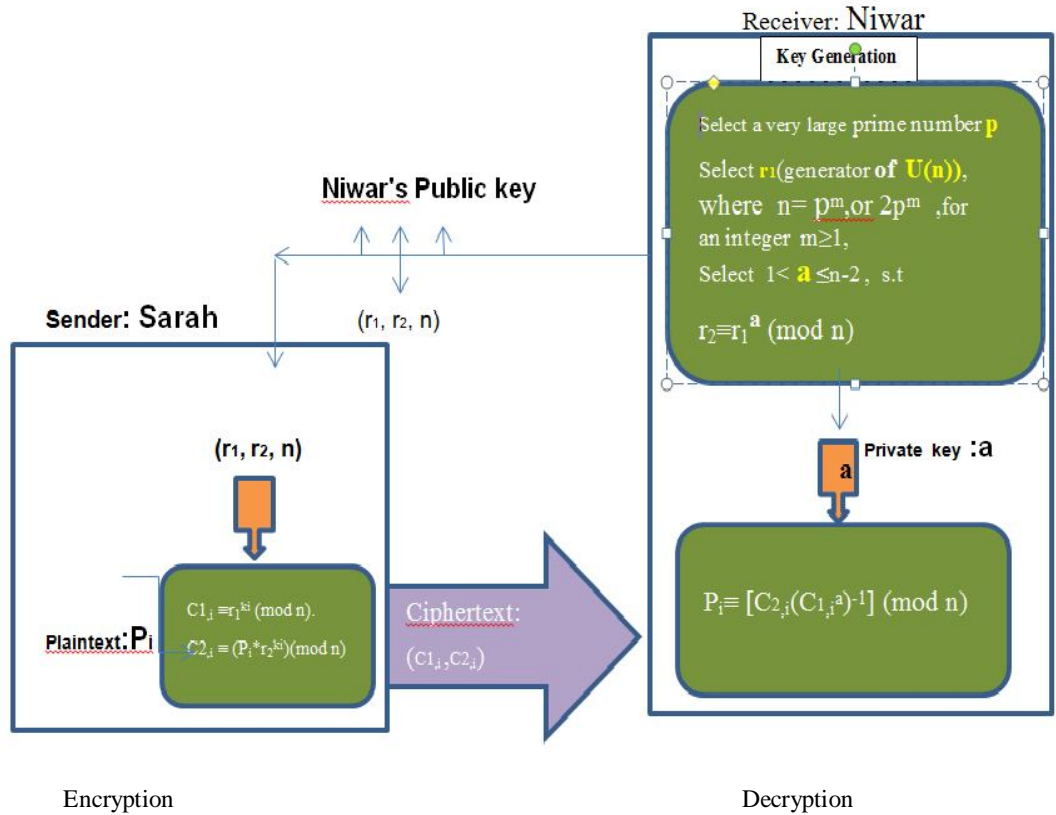

      Encryption                          Decryption

( Figure 01)

Fig. 1 A simple graph to show the procedure of ElGamal Cryptosystem over U(n).

TABLE 1

Letters Numerical equivalents

| Letter | A | B | C | D | E | F | G | H | I | J | K | L | M | N | O | P | Q | R | S | T | U | V | W | X | Y | Z |
|---|---|---|---|---|---|---|---|---|---|---|---|---|---|---|---|---|---|---|---|---|---|---|---|---|---|---|
| Numerical Equivalent P: | 0 | 1 | 2 | 3 | 4 | 5 | 6 | 7 | 8 | 9 | 10 | 11 | 12 | 13 | 14 | 15 | 16 | 17 | 18 | 19 | 20 | 21 | 22 | 23 | 24 | 25 |

TABEL 02
Computing $C_{2,i}$ for each block $P_i$ in Example 1.

| The Plaintext $P_i$: | $C_{2,i} \equiv P_i*(r_2)^k \pmod{n} \equiv P_i * 23^5 \pmod{29} \equiv P_i * 25 \pmod{29}$ | Ciphertext: $E(P_i)=(C_{1,i},C_{2,i})$ |
|---|---|---|
| P1: 08 | $C_{2,1} \equiv 08*25 \pmod{29} = 26$ | (11,26) |
| P2: 11 | $C_{2,2} \equiv 11*25 \pmod{29} = 14$ | (11,14) |
| P3: 08 | $C_{2,3} \equiv 08*25 \pmod{29} = 26$ | (11,26) |
| P4: 10 | $C_{2,4} \equiv 10*25 \pmod{29} = 18$ | (11,18) |
| P5: 04 | $C_{2,5} \equiv 04*25 \pmod{29} = 13$ | (11,13) |
| P6: 12 | $C_{2,6} \equiv 12*25 \pmod{29} = 10$ | (11,10) |
| P7: 00 | $C_{2,7} \equiv 00*25 \pmod{29} = 0$ | (11,0) |
| P8: 19 | $C_{2,8} \equiv 19*25 \pmod{29} = 11$ | (11,11) |
| P9: 07 | $C_{2,9} \equiv 07*25 \pmod{29} = 1$ | (11,1) |

TABEL 03
Computing $P_i$ for each $C_{2,i}$ in Example 1

| CIPHERTEXT: $E(P_I)=(C_{1,I},C_{2,I})$ | PLAINTEXT: $P_I \equiv (C_{2,I})((C_{1,I})^A)^{-1} \pmod{P} \equiv (C_{2,I})((11)^4)^{-1} \pmod{29} \equiv (C_{2,I})(7) \pmod{29}$ |
|---|---|
| $(C_{1,1},C_{2,1})=(11,26)$ | $P_1 \equiv 26*7 \pmod{29} = 08$ |
| $(C_{1,2},C_{2,2})=(11,14)$ | $P_2 \equiv 14*7 \pmod{29} = 11$ |
| $(C_{1,3},C_{2,3})=(11,26)$ | $P_3 \equiv 26*7 \pmod{29} = 08$ |
| $(C_{1,4},C_{2,4})=(11,18)$ | $P_4 \equiv 18*7 \pmod{29} = 10$ |
| $(C_{1,5},C_{2,5})=(11,13)$ | $P_5 \equiv 13*7 \pmod{29} = 04$ |
| $(C_{1,6},C_{2,6})=(11,10)$ | $P_6 \equiv 10*7 \pmod{29} = 12$ |
| $(C_{1,7},C_{2,7})=(11,0)$ | $P_7 \equiv 0*7 \pmod{29} = 00$ |
| $(C_{1,8},C_{2,8})=(11,11)$ | $P_8 \equiv 11*7 \pmod{29} = 19$ |
| $(C_{1,9},C_{2,9})=(11,01)$ | $P_9 \equiv 01*7 \pmod{29} = 07$ |

## II. Conclusion

From the modification of discrete logarithm problem based cryptosystem, it is clear to see that the discrete logarithm problem based cryptosystem (especially over ElGamal Cryptosystem ) works easily over another type of the groups (other than the finite cyclic group ) which is the abelian group U(n) where  n= $p^m$, or $2p^m$ , for  an integer  m≥1. And this new study might maintain equivalent (or better) security with the original ElGamal cryptosystem over $Z^*_p$ because in the original ElGamal cryptosystem, U(n=$p^1$) was used and given a really strong security as Taher ElGamal claimed and proved. But in this new study I modify the ElGamal cryptosystem using  n= $p^m$, or n= $2p^m$ , for  an integer  m≥1. And it absolutely gives a better security if we apply and test it because finding the private key depends on finding all the possible solutions for the discrete logarithm problem, which is impossible over bigger groups like U(n) where n= $p^m$, or n= $2p^m$ , for  an integer  m≥1 .